\documentclass[useAMS,usenatbib,usegraphicx]{mn2e}
\usepackage{ gensymb }

\usepackage{graphicx}
\usepackage[fleqn]{amsmath}
\usepackage{amssymb}
\usepackage{bm}
\usepackage{verbatim}
\usepackage{color}
\usepackage{ulem}

\newcommand{\dd}{{\rm d}}

\def\ga{\,\,\raise0.14em\hbox{$>$}\kern-0.76em\lower0.28em\hbox
{$\sim$}\,\,}
\def\la{\,\,\raise0.14em\hbox{$<$}\kern-0.76em\lower0.28em\hbox
{$\sim$}\,\,}

\title[MRI in neutron star mergers]
{Magnetorotational instability in neutron star mergers: impact of neutrinos}

\author[Guilet et al.]{J\'er\^ome Guilet$^{1,2,3}$, Andreas Bauswein$^{4}$, Oliver Just$^{1,2}$ \& Hans-Thomas Janka$^{1}$ \\
$^1$ Max-Planck-Institut f\"ur Astrophysik, Karl-Schwarzschild-Str. 1, D-85748 Garching, Germany \\ 
$^2$ Max Planck/Princeton Center for Plasma Physics \\
$^3$ Laboratoire AIM, CEA/DSM-CNRS-Universit{\'e} Paris 7, Irfu/Service d'Astrophysique, CEA-Saclay, 91191 Gif-sur-Yvette, France \\
$^4$ Heidelberger Institut f\"ur Theoretische Studien, Schloss-Wolfsbrunnenweg 35, D-69118 Heidelberg, Germany}

\begin{document}

\maketitle

\label{firstpage}

\begin{abstract}
The merger of two neutron stars may give birth to a long-lived hypermassive neutron star. If it harbours a strong magnetic field of magnetar strength, its spin-down could explain several features of short gamma-ray burst afterglows. The magnetorotational instability (MRI) has been proposed as a mechanism to amplify the magnetic field to the required strength. Previous studies have, however, neglected neutrinos, which may have an important impact on the MRI by inducing a viscosity and drag. We investigate the impact of these neutrino effects on the linear growth of the MRI by applying a local stability analysis to snapshots of a neutron star merger simulation. We find that neutrinos have a significant impact inside the hypermassive neutron star, but have at most a marginal effect in the torus surrounding it. Inside the hypermassive neutron star, the MRI grows in different regimes depending on the radius and on the initial magnetic field strength. For magnetic fields weaker than $10^{13}-10^{14}\,{\rm G}$, the growth rate of the MRI is significantly reduced due to the presence of neutrinos. We conclude that neutrinos should be taken into account when studying the growth of the MRI from realistic initial magnetic fields. Current numerical simulations, which neglect neutrino viscosity, are only consistent, i.e. in the adopted ideal regime, if they start from artificially strong initial magnetic fields above $\sim10^{14}\,{\rm G}$. One should be careful when extrapolating these results to lower initial magnetic fields, where the MRI growth is strongly affected by neutrino viscosity or drag.
\end{abstract}


\section{Introduction}
Neutron star (NS) mergers are the leading candidates to power short gamma-ray bursts
\citep[short GRBs,][]{Paczynski1986,Eichler1989,barthelmy05} and are one of the most promising targets for the new
generation of gravitational wave detectors (LIGO, VIRGO) \citep{abadie10}.  The nature of the
compact object formed during the merger and powering short gamma-ray bursts is debated. It could be
either a black hole (BH) or a rapidly rotating, massive, strongly magnetized NS
\citep{usov92,metzger11}. Such a supermassive magnetar is invoked to explain the extended emission and the
X-ray light curves observed in a number of short GRBs
\citep{metzger08,bucciantini12,rowlinson13,gompertz13,gompertz14,lu15} and could also lead to late
UV, visible and infrared emission \citep{fan13,metzger14}.

The evolution and impact of the magnetic field in NS mergers have been explored through
numerical simulations
\citep{liu08,anderson08,rezzolla11,giacomazzo11,giacomazzo15,kiuchi14,kiuchi15,palenzuela15,kawamura16}. An
amplification of the magnetic field to magnetar strength in the hypermassive NS
(HMNS) has been suggested to occur due to several potential amplification mechanisms. First,
the shear layer that forms where the surfaces of the NSs touch is subject to the
Kelvin-Helmholtz instability and the turbulence it generates can lead to a small-scale dynamo
\citep[e.g.][]{price06,obergaulinger10,kiuchi15}. Second, the magnetorotational instability
\citep[e.g.][]{balbus91} has also been argued to be at work as suggested by numerical
simulations \citep{duez06,siegel13,kiuchi14}. The wavelength of the MRI, which is proportional to the
magnetic field, can only be resolved by current global numerical simulations if the
initial magnetic field is already extremely strong ($10^{15}\,{\rm G}$ and higher). Observed binary
pulsars have a surface dipole magnetic-field strength in the range of $10^{10}\,{\rm G}$ to $10^{12}\,{\rm G}$
\citep{lorimer08}. It therefore remains to be investigated whether the growth of the MRI would be
significantly different for the weaker initial fields suggested by these observations. In this
paper, we investigate the so-far neglected impact of neutrinos on the MRI in NS
mergers. We find that it is significant for realistic initial magnetic fields and that the MRI then
grows in regimes different from the high magnetic field case considered so far.

In Section~\ref{sec:numerical_model}, we describe the numerical model of a NS merger
that we use. In Section~\ref{sec:MRI}, analytical predictions for the growth of the MRI in the
presence of neutrino radiation are applied to this numerical model. In Section~\ref{sec:KH}, the
impact of neutrinos on the Kelvin-Helmholtz instability is discussed. Finally, in
Section~\ref{sec:conclusion}, we discuss and conclude on the consequences of our results for neutron
star mergers.

\section{Numerical model}\label{sec:numerical_model}
\begin{figure}
\centering
 \includegraphics[width=\columnwidth]{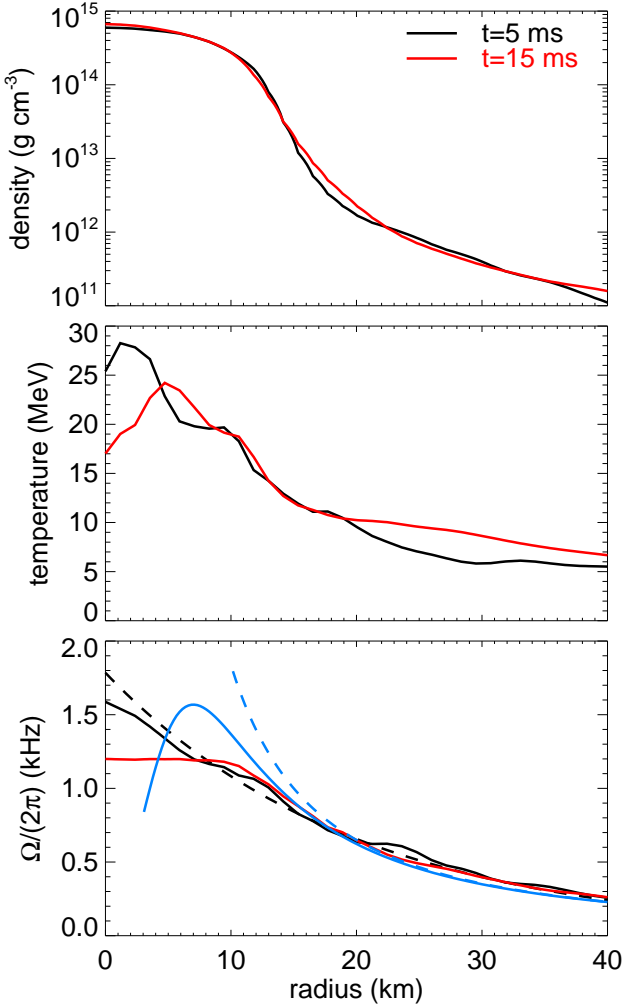}
 \caption{Azimuthally averaged radial profiles in two snapshots taken $5\,{\rm ms}$ (black lines) and $15\,{\rm ms}$ (red lines) after the collision of the NSs. Shown is the density (top), temperature (middle) and rotation frequency (bottom). The exponential fit to the angular frequency profile at the early time is shown with a black dashed line (equation~\ref{eq:exponential_fit}). The blue solid line in the bottom panel shows the rotation frequency profile defined by equation~(\ref{eq:hanauske_profile}), while the dashed blue line shows the power law profile $\Omega=\Omega_1\left(r/r_1\right)^{-3/2}$ for comparison. The rotation profile of equation~(\ref{eq:hanauske_profile}) approximates the behavior of the models with the highest maximum of $\Omega$ in Figure~11 of \citet{hanauske16} (their models APR4-M125 and APR4-M135).}
             \label{fig:profiles}%
\end{figure}

In order to estimate the physical quantities relevant for the growth of the MRI, we use the result
of a numerical simulation of a NS merger. The computations were performed
  with a relativistic smoothed particle hydrodynamics code, which imposes the conformal flatness
  condition on the spatial metric for solving the Einstein equations
  \citep{isenberg80,wilson96}. Details of the numerical and physical model can be found
  in~\citet{oechslin07,bauswein10}. For this study we employed a temperature-dependent, microphysical
  high-density equation of state, specifically we used the DD2 model \citep{hempel10}, which is based
  on the relativistic mean-field parameterization proposed by~\citet{typel10}. This equation of
  state is moderately stiff and results in a NS radius of 13.2~km for a star with a
  gravitational mass of 1.35~$M_{\odot}$. We focus on a simulation of a binary with two stars of
  1.35~$M_{\odot}$, which can be considered to be representative with respect to the observed
  distribution of masses in NS binaries \citep[see, e.g., the compilation
  in][]{lattimer12}. This binary setup leads to the formation of a rapidly rotating NS merger remnant, which is stable until the end of the simulation (several ten milliseconds  after merging). The formation of a stable stellar remnant is expected for a large variety of equation of state models
  in the total binary mass range of about 2.4~$M_{\odot}$ to 3.0~$M_{\odot}$, which is
  representative for the observed binaries \citep{bauswein13}. From the hydrodynamical data of the
  postmerger object we extract radial profiles of various quantities by averaging over the
azimuthal direction and we restrict our analysis to the equatorial plane for
simplicity. Figure~\ref{fig:profiles} shows the radial profiles at the two times we consider in the
analysis : $5\,{\rm ms}$ and $15\,{\rm ms}$ after the collision of the NSs. At the early time, the
HMNS is not yet axisymmetric such that this procedure only gives a rough estimate of the conditions
locally felt by the MRI. This should be less problematic at the later time we consider. Numerical
diffusion inherent to any numerical simulation probably leads to significant uncertainties in
particular in two aspects. First, the flattening of the rotation profile in the central $5-10\,{\rm km}$ of the HMNS observed at the later time may be accelerated by numerical viscosity. Second, the temperature is quite uncertain as it is determined by the conversion of
kinetic to thermal energy, which is also sensitive to the numerical viscosity. In order to estimate
how this uncertainty impacts our results we also considered temperature profiles $20\%$ higher or
lower than the output of the simulation (see Section~\ref{sec:drag_MRI}). This has a
significant quantitative impact on our results but we expect the qualitative picture to remain
robust. Similarly, we anticipate that models with different binary masses or
  a different equation of state lead to quantitative differences but leave the qualitative results
  unchanged. Moreover, we caution that a merger remnant initially shows a complex velocity field, which significantly deviates from axisymmetry, while we consider an azimuthally averaged angular velocity and neglect gauge effects as well \citep{kastaun15}.

The somewhat noisy rotation profile at the early time is problematic for the analysis because the
results depend critically on the radial derivative of the angular frequency. To avoid this problem,
we replace the numerical profile by the following fit of the angular frequency $\Omega$ as a function of radius $r$ (dashed black line in
Figure~\ref{fig:profiles}): 
\begin{equation}
\Omega = \Omega_0\exp\left(-r/r_0 \right),
	\label{eq:exponential_fit}
\end{equation}
with $\Omega_0 = 1.12\times 10^3 \,{\rm s^{-1}}$ and $r_0=20\,{\rm km}$. The relation of these angular momentum profiles to results in the literature and the influence on our analysis will be discussed in Section~\ref{sec:rotation_profile}.

\section{MRI growth in the presence of neutrino radiation}\label{sec:MRI}

In this section, we apply analytical estimates obtained in \citet{guilet15} for the growth of the
MRI in the presence of neutrino radiation. These analytical results were obtained under a number of
assumptions, which we quickly summarize \citep[see][for a more detailed discussion]{guilet15}. We
use a local dispersion relation, which is valid if the wavelength of the modes is much shorter than the typical
lengthscales of the object. This is well justified for moderate magnetic fields (see
Section~\ref{sec:sigma_lambda}). For simplicity, the magnetic field is
assumed to be perpendicular to the equatorial plane, which is the most classical and favourable configuration for MRI
growth. Only axisymmetric modes are considered as these are the most unstable ones under this
magnetic field geometry. The resistivity is neglected because it is expected to be extremely
small \citep{thompson93}. We also neglect the impact of the (presumably stable) stratification connected to entropy and
composition gradients. A stable stratification could in principle have a significantly stabilising
influence on the MRI, but neutrino diffusion is expected to alleviate this impact
\citep{menou04,masada07,guilet15b}. We chose to neglect it for simplicity because the
composition and entropy gradients are not perfectly axisymmetric and quite
uncertain (see Section~\ref{sec:numerical_model}), which hinders a robust conclusion on their impact on the MRI. Finally, all the
calculations are non-relativistic because no relativistic linear analysis of the MRI including
neutrinos has been performed so far. This is most likely not a major limitation, because general
relativistic analyses in ideal MHD suggest that the MRI is not critically changed with respect to a
Newtonian approximation even for a Kerr metric \citep{araya-gochez02,gammie04,yokosawa05}.

\subsection{MRI with neutrino viscosity (diffusive regime)}
	\label{sec:viscous_MRI}
\begin{figure}
\centering
  \includegraphics[width=\columnwidth]{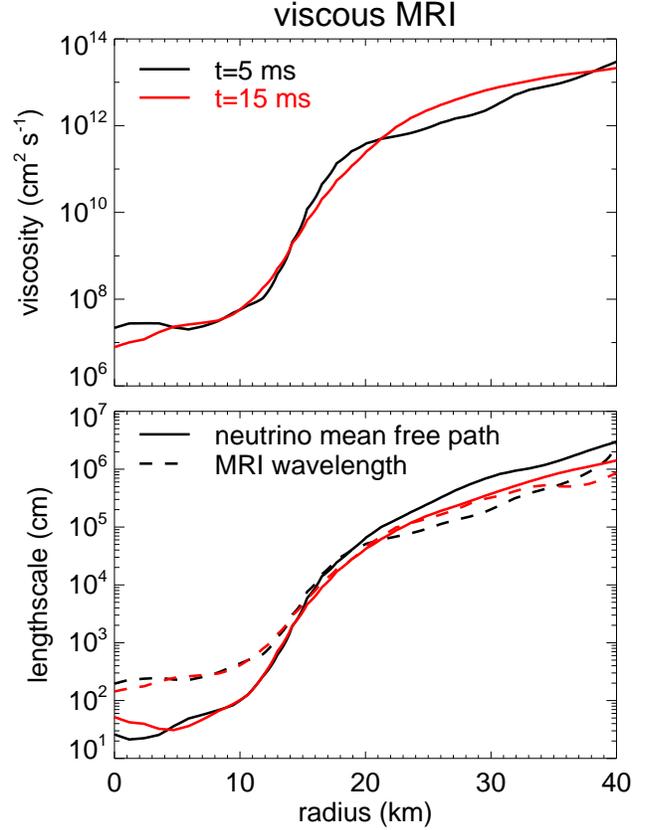}
  \caption{Top panel: radial profile of neutrino viscosity in the equatorial
    plane. Bottom panel: wavelength of the most unstable MRI mode in the viscous regime computed using equation~\ref{eq:k_viscous} (dashed line)
    compared to the mean free path of heavy lepton neutrinos (solid lines). The two snapshots are
    shown in black ($5\,{\rm ms}$ after the collision of the NSs) and red ($15\,{\rm ms}$ after the collision of the NSs).}
              \label{fig:neutrino_viscosity}%
\end{figure}	

On lengthscales longer than the mean free path of neutrinos, neutrino diffusion induces a
viscosity. We estimate it by using the approximate analytical expression obtained by \citet{keil96}
\begin{equation}
\nu = 1.2\times 10^{10} \left(\frac{T}{10\,{\rm MeV}}\right)^2\left(\frac{\rho}{10^{13}\,{\rm g\,cm^{-3}}}\right)^{-2} \,{\rm cm^2\,s^{-1}},
	\label{eq:neutrino_viscosity_approx}
\end{equation}
where $\nu$ is the kinematic viscosity, $\rho$ is the density and $T$ the
  temperature. This expression was successfully compared to a direct calculation of the neutrino
viscosity using the neutrino distribution from a numerical simulation of protoneutron star formation
with elaborate neutrino transport \citep{guilet15}. The viscosity we estimate from the simulation
outputs is shown in the upper panel of Figure~\ref{fig:neutrino_viscosity}: it increases strongly
with radius from $\sim 10^7\,{\rm cm^2 s^{-1}}$ in the inner core of the HMNS to
$\sim 10^{11}-10^{12}\,{\rm cm^2 s^{-1}}$ near the surface of the HMNS\footnote{We define the surface of the HMNS as the location beyond which the rotational support exceeds the pressure support. It is located at a radius of about $20\,{\rm km}$.}.

The effect of viscosity on the linear growth of the MRI is controlled by the viscous Elsasser number
$E_\nu \equiv \frac{v_{A}^2}{\nu\Omega}$ \citep[e.g.][]{pessah08,longaretti10,guilet15}, where
$v_A \equiv B/\sqrt{4\pi\rho}$ is the Alfv\'en speed with the magnetic field strength $B$. For
$E_\nu < 1$, viscosity affects significantly the growth of the MRI: as a result the growth rate is
decreased, and the wavelength of the most unstable mode becomes longer. Typical conditions inside
the HMNS lead to the following estimate of the viscous Elsasser number at a radius of 
$\sim 5-10\,{\rm km}$
\begin{eqnarray}
E_\nu &\sim& 8\times 10^{-4} \left(\frac{B}{10^{12}\,{\rm G}} \right)^{2}\left(\frac{\rho}{4\times10^{14}\,{\rm g\,cm^{-3}}} \right)^{-1}  \nonumber \\
&&\times \left(\frac{\Omega}{6000\,{\rm s^{-1}}} \right)^{-1} \left(\frac{\nu}{4\times10^{7}\,{\rm cm^2\,s^{-1}}} \right)^{-1}.
\end{eqnarray}
Viscosity therefore has a large impact on the linear growth of the MRI, unless the
magnetic field is significantly stronger than the surface field of normal pulsars or another
  mechanism significantly amplifies an initially weak field. The critical strength of the magnetic
field below which viscous effects become important (at which $E_\nu=1$) is
\begin{eqnarray}
B_{\rm visc} &=& \sqrt{4\pi\rho \nu\Omega}  \nonumber	\label{eq:Bvisc} \\
&=& 3.5\times10^{13}\left(\frac{\rho}{4\times10^{14}\,{\rm g\,cm^{-3}}} \right)^{1/2} \left(\frac{\nu}{4\times10^{7}\,{\rm cm^2\,s^{-1}}} \right)^{1/2} \nonumber \\
&&\times \left(\frac{\Omega}{6000\,{\rm s^{-1}}} \right)^{1/2} \,{\rm G}.
\label{eq:Bvisc2}
\end{eqnarray}

The dispersion relation governing the growth of the MRI in the presence of viscosity reads
\citep{lesur07,pessah08,masada08} 
\begin{equation}
\left\lbrack\left(\sigma+k^2\nu\right)\sigma +k^2v_A^2\right\rbrack^2  + \kappa^2 \left(\sigma^2 + k^2v_A^2 \right) - 4\Omega^2k^2v_A^2  = 0,
	\label{eq:dispersion_viscous}
\end{equation}
where $\sigma$ is the growth rate, $k$ the wavenumber, and $\kappa$ the epicyclic frequency defined by $\kappa^2 \equiv \frac{1}{r^3} \frac{\dd (r^4 \Omega^2)}{\dd r}$.

In the asymptotic limit $E_\nu \ll 1$, the growth rate and wavelength of the most unstable mode can then be expressed as \citep{pessah08,masada08,masada12}
\begin{eqnarray}
\sigma &=&  \left(\frac{qE_\nu}{\tilde\kappa} \right)^{1/2} \Omega = 90  \left(\frac{B}{10^{12}\,{\rm G}} \right)\left(\frac{\rho}{4\times10^{14}\,{\rm g\,cm^{-3}}} \right)^{-1/2} \nonumber \\
&& \times \left(\frac{\nu}{4\times10^{7}\,{\rm cm^2\,s^{-1}}} \right)^{-1/2} \left(\frac{\Omega}{6000\,{\rm s^{-1}}} \right)^{1/2} \,{\rm s^{-1}},
	\label{eq:sigma_viscous}
\end{eqnarray} 
\begin{eqnarray}
\lambda &=& 2\pi\left(\frac{\nu}{\kappa} \right)^{1/2} \nonumber \\
&=& 390  \left(\frac{\kappa}{6000\,{\rm s^{-1}}} \right)^{-1/2}\left(\frac{\nu}{4\times10^{7}\,{\rm cm^2\,s^{-1}}} \right)^{1/2} \,{\rm cm},
	\label{eq:k_viscous}
\end{eqnarray}
where $\tilde\kappa \equiv \kappa/\Omega = \sqrt{2(2-q)}$ is the dimensionless epicyclic frequency, and $q \equiv - \dd \log\Omega/\dd\log r=0.5$ in the numerical estimates. In contrast to the inviscid case, the wavelength of the fastest growing mode is independent of the magnetic field strength and it is longer than it would be without viscosity. The growth rate on the other hand is reduced and is proportional to the magnetic field strength. The MRI therefore requires a minimum magnetic field strength in order to grow on a given  timescale. The minimum magnetic field necessary for the MRI to grow at a minimum growth rate $\sigma_{\rm min}$ is \citep{guilet15}
\begin{eqnarray}
B_{\rm min} &=& \left(\frac{4\pi\rho \tilde\kappa \nu}{q\Omega} \right)^{1/2}\sigma_{\rm min},  \label{eq:viscous_Bmin}	\\
&=& 10^{12} \left(\frac{\sigma_{\rm min}}{100\,{\rm s^{-1}}} \right) \left(\frac{\rho}{4\times10^{14}\,{\rm g\,cm^{-3}}} \right)^{1/2} \nonumber \\
&& \left(\frac{\nu}{4\times10^{7}\,{\rm cm^2\,s^{-1}}} \right)^{1/2} \left(\frac{\Omega}{6000\,{\rm s^{-1}}} \right)^{-1/2} {\rm G}.   \label{eq:viscous_Bmin2}
\end{eqnarray}

The viscous regime is valid if the wavelength of the fastest growing mode is larger than the neutrino mean free path. Figure~\ref{fig:neutrino_viscosity} (lower panel) shows that this is indeed the case inside the HMNS but not in the surrounding torus. In that figure, the mean free path of heavy lepton neutrinos (which give the most restrictive constraint among the neutrino species) is estimated using the scaling relation $l_\nu = 10^4 (\rho/10^{13}\,{\rm g\,cm^{-3}})^{-1}(T/10\,{\rm MeV})^{-2} \,{\rm cm}$ \citep{guilet15}.


\subsection{MRI with neutrino drag}
	\label{sec:drag_MRI}
\begin{figure}
\centering
  \includegraphics[width=\columnwidth]{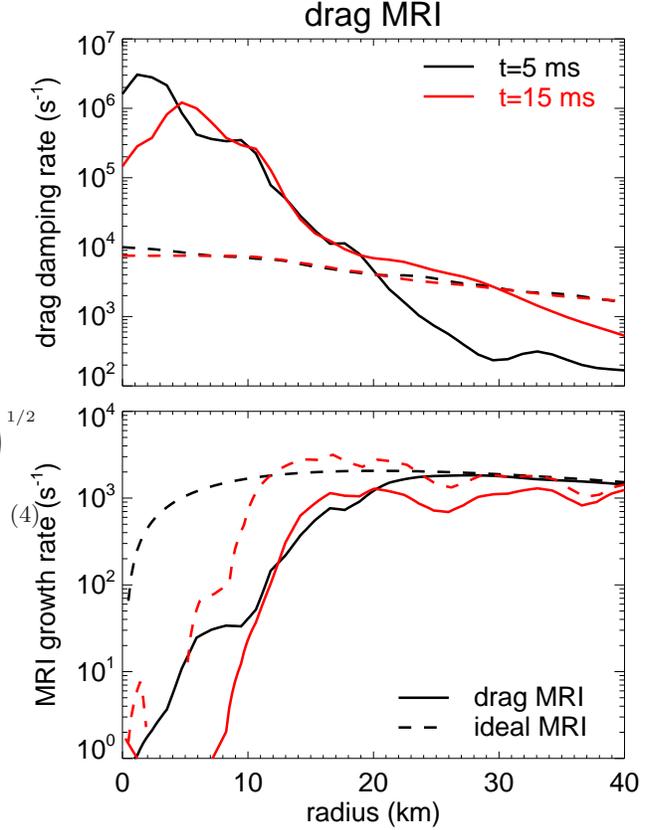}
 \caption{Top panel: Radial profile of the neutrino drag damping rate (solid lines) compared to the angular frequency~(dashed lines). Bottom panel: growth rate of the MRI including (solid lines) or neglecting (dashed lines) the impact of neutrinos. The black and red lines correspond to the early and late times, respectively.}
               \label{fig:neutrino_drag}%
\end{figure}

At wavelengths shorter than their mean free path, neutrinos induce a drag force of the form $-\Gamma {\bf v}$, where $\Gamma$ is a damping rate depending on the neutrino distribution and opacity \citep{jedamzik98,thompson05,guilet15}. We compute the neutrino drag damping rate from the simulation outputs using the scaling obtained by \citet{guilet15} 
\begin{equation}
\Gamma = 6\times 10^3\,(T/10\,{\rm MeV})^6\,{\rm s^{-1}}.
	\label{eq:gamma_scaling}
\end{equation}

\begin{figure}
\centering
  \includegraphics[width=\columnwidth]{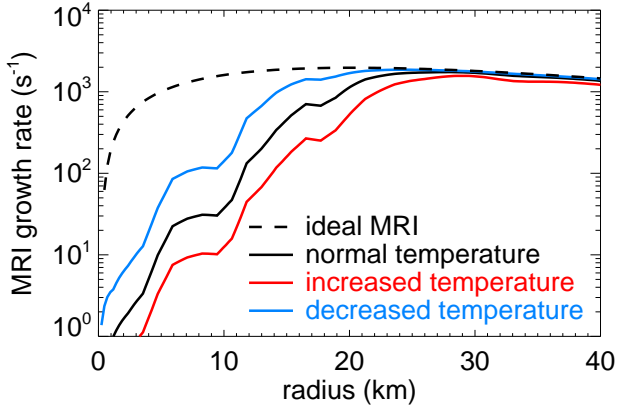}
 \caption{Growth rate of the MRI in the drag regime at early time ($t=5\,{\rm ms}$) when the temperature is scaled up (red line) or down (blue line) by $20\%$. The temperature uncertainty has a significant quantitative impact but leaves the qualitative conclusions unaffected.}
              \label{fig:temperature_dependence}%
\end{figure}

The dispersion relation governing the growth of the MRI in the presence of neutrino drag is \citep{guilet15} 
\begin{equation}
\left\lbrack\left(\sigma+\Gamma\right)\sigma +k^2v_A^2\right\rbrack^2  + \kappa^2 \left(\sigma^2 + k^2v_A^2 \right) - 4\Omega^2k^2v_A^2  = 0.
	\label{eq:dispersion_drag}
\end{equation}
The neutrino drag impacts significantly the MRI when $\Gamma\gtrsim\Omega$. Figure~\ref{fig:neutrino_drag} shows that this is the case at all radii inside the HMNS ($r\lesssim20\,{\rm km}$) but not in the surrounding torus ($r\gtrsim20\,{\rm km}$). In the asymptotic limit $\Gamma \gg \Omega$, the growth rate and wavenumber of the fastest growing mode are \citep{guilet15}
\begin{equation}
\sigma = \frac{q}{2}\frac{\Omega^2}{\Gamma},
	\label{eq:sigma_drag}
\end{equation}
and
\begin{equation}
k = \sqrt{q/2}\frac{\Omega}{v_A},
	\label{eq:k_drag}
\end{equation}
respectively. Compared to the ideal MHD case, the maximum growth rate is reduced by a factor $\Gamma/\Omega$ while the wavelength of the fastest growing mode is not drastically changed. Importantly, and contrary to the viscous regime, the growth rate is independent of the magnetic field strength. The growth rate taking into account the neutrino drag is compared to the ideal MHD case in the bottom panel of Figure~\ref{fig:neutrino_drag}. As expected from the above arguments, the growth of the MRI is significantly slowed down by the neutrino drag inside the HMNS. Because of the steep dependence of the neutrino drag damping rate on temperature, the growth rate in this regime is somewhat uncertain. This is illustrated by Figure~\ref{fig:temperature_dependence} where the impact of varying the temperature profile by $20\%$ is shown to have quantitatively significant consequences but without changing the qualitative picture that neutrino drag reduces the MRI growth rate significantly inside the HMNS but impacts it at most marginally in the surrounding torus.

The neutrino drag regime is valid if the wavelength of the fastest growing MRI mode is shorter than the neutrino mean free path. Using equation~(\ref{eq:k_drag}), this condition can be expressed as an upper limit on the magnetic field strength,
\begin{equation}
B<\sqrt{\rho q/2\pi} \Omega l_\nu.
	\label{eq:B_consistency_drag}
\end{equation}
This critical magnetic field is shown as a dashed line in Figures~\ref{fig:MRI_regimes} and \ref{fig:MRI_regimes_2} with the mean free path of electron neutrinos estimated as $l_\nu = 2\times10^3 (\rho/10^{13}\,{\rm g\,cm^{-3}})^{-1}(T/10\,{\rm MeV})^{-2} \,{\rm cm}$.

\subsection{Growth rate and wavelength}
	\label{sec:sigma_lambda}

\begin{figure}
\centering
 \includegraphics[width=\columnwidth]{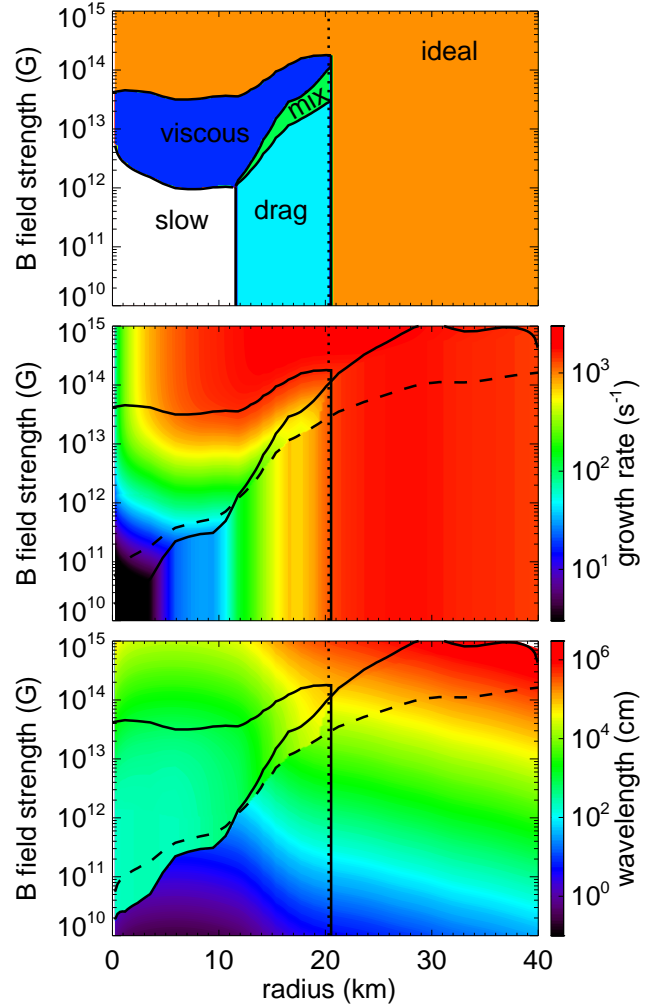}
 \caption{Properties of the MRI in a snapshot $5\,{\rm ms}$ after the collision of the NSs. Top panel: Different
   regimes of MRI growth as a function of radius and magnetic field strength. The regime labelled
   slow corresponds to growth timescales longer than $10\,{\rm ms}$. Middle panel: growth rate of
   the MRI. Bottom panel: wavelength of the most unstable MRI mode. The vertical dotted lines
   show the boundary between the pressure supported HMNS and the rotation supported
   torus surrounding it. The lines separating the different regimes are given by
   equation~\ref{eq:Bvisc} (giving the limit between viscous and ideal regimes),
   equation~\ref{eq:B_consistency_drag} (delineating the boundary between "drag" and "mixed" regimes,
   and shown as a dashed line in middle and bottom panels) and equation~\ref{eq:B_visc-drag}
   (solid line going from bottom left to top right). The vertical solid line separating the drag and ideal regimes shows the radius
   at which $\Gamma=\Omega$.}
              \label{fig:MRI_regimes}%
\end{figure}

\begin{figure}
\centering
 \includegraphics[width=\columnwidth]{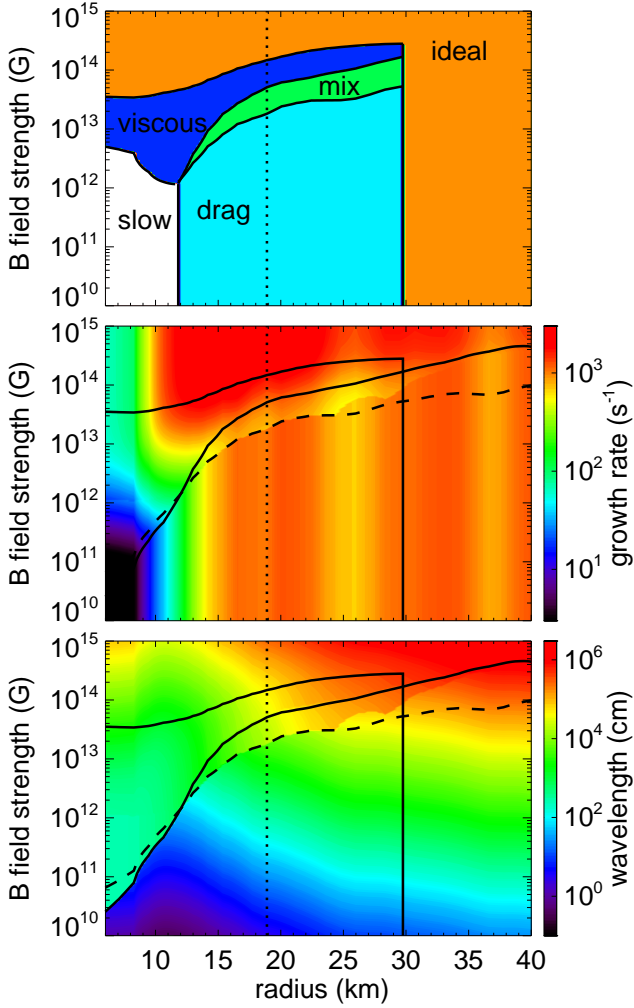}
 \caption{Same as Figure~\ref{fig:MRI_regimes} but at a later time, $15\,{\rm ms}$ after the collision of the NSs. The range of radii has been limited to $r>6\,{\rm km}$ to exclude the MRI-stable inner core in solid-body rotation.}
              \label{fig:MRI_regimes_2}%
\end{figure}	

In this section we determine the relevant regime of MRI growth as a function of radius and magnetic
field strength. This then allows us to compute the growth rate and wavelength of the most unstable
mode (Figures~\ref{fig:MRI_regimes} and \ref{fig:MRI_regimes_2}). At any radius, both the viscous
and the drag regimes can in principle be relevant, and one has to compare the growth rate in each
regime in order to obtain the fastest growing mode. Since the growth rate in the viscous regime is
proportional to the magnetic field strength, the growth is faster in the viscous regime (dark blue in top panel of Figure~\ref{fig:MRI_regimes} and \ref{fig:MRI_regimes_2}) than in the drag regime (light blue) above a critical magnetic field strength, which can be obtained
by combining equations~(\ref{eq:sigma_viscous}) and (\ref{eq:sigma_drag})
\begin{equation}
B_{\rm visc-drag} = \sqrt{q\pi\rho\nu\kappa}\frac{\Omega}{\Gamma}.
	\label{eq:B_visc-drag}
\end{equation}
This expression is valid when $E_\nu\ll 1$ and $\Gamma/\Omega\gg 1$ and the result is shown by the
solid line going from the bottom left to upper right in the middle and bottom panels of
Figures~\ref{fig:MRI_regimes} and \ref{fig:MRI_regimes_2}. There is also a mixed regime of MRI
growth (green) where some neutrino species are in the viscous regime while some are in the drag
regime. There is no analytical description of this regime so far and since it is quite limited in
parameter space, we do not attempt to derive one. Instead, we adopt the practical and very rough
approach of using the average of the values obtained in the viscous and drag regimes for the growth rate and wavelength of the most unstable mode.

The growth rate and wavelength of the fastest growing modes are obtained by solving numerically the
dispersion relation in each relevant regime (equations~\ref{eq:dispersion_viscous} and
\ref{eq:dispersion_drag}). The result is shown as a function of radius and magnetic field strength
in Figure~\ref{fig:MRI_regimes} for the early snapshot ($5\,{\rm ms}$ after the collision of the NSs) and
Figure~\ref{fig:MRI_regimes_2} for the late snapshot ($15\,{\rm ms}$). The MRI can grow unabated on
timescales of $1\,{\rm ms}$ or shorter for magnetic fields stronger than $\sim10^{14}\,{\rm G}$
inside the HMNS and for any magnetic field in the surrounding torus. Inside the HMNS and for weaker
magnetic fields, the MRI growth is slowed down by neutrinos in either of the two regimes: viscous or
drag. The growth timescale is longer than $10\,{\rm ms}$ for densities above
  roughly $10^{14}\,{\rm g\,cm^{-3}}$ and magnetic fields smaller than $10^{12}\,{\rm
  G}$. In the viscous regime the wavelength of the fastest growing mode is of the order of a few meters,
but can be much shorter in the drag regime ($<1\,{\rm cm}$). These short lengthscales highlight the
difficulty of resolving the MRI in a global numerical simulation. Our analysis also stresses the fact that if one avoids this problem by starting with a strong initial magnetic field, the MRI then grows in a very different physical regime.

\subsection{Dependence on the angular velocity profile}
	\label{sec:rotation_profile}
	
\begin{figure}
\centering
 \includegraphics[width=\columnwidth]{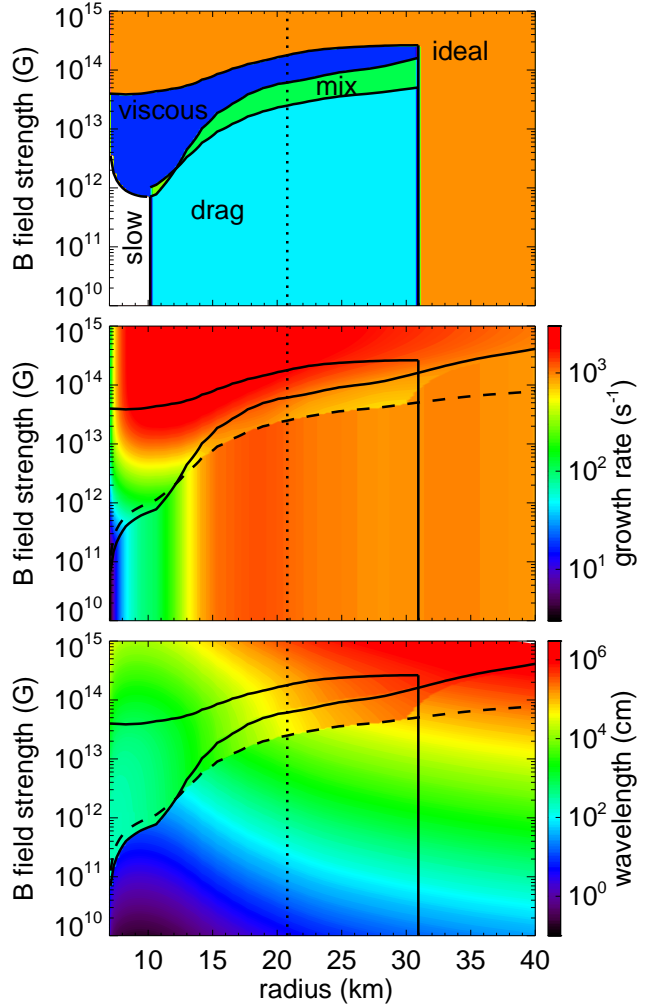}
 \caption{Same as Figure~\ref{fig:MRI_regimes_2} but with the rotation profile as defined by equation~(\ref{eq:hanauske_profile}). The range of radii has been limited to $r>7\,{\rm km}$ to exclude the MRI-stable inner core where $\Omega$ increases with the radius.}
              \label{fig:MRI_regimes_3}%
\end{figure}	

It is important to compare our rotation profiles to the results published for simulations performed with grid-based codes \citep{shibata05,shibata06_ns-merger,kastaun15,endrizzi16,kastaun16a,kastaun16b,hanauske16}. Overall, these simulations suggest a rotation frequency profile that has a maximum at the center at early times (for a few milliseconds after the merger). This profile then evolves to one with a slowly rotating core, where the rotation frequency increases with radius up to a maximum (in the range $\Omega\sim(6-10)\times10^3\,{\rm s^{-1}}$ at a radius of $r\sim5-15\,{\rm km}$ depending on the simulations) and decreases outward. Our early-time profile is qualitatively consistent with these results at early times. Our late-time profile, on the other hand, is flat in the center with a maximum value of $\Omega\simeq7.5\times10^3{\rm s^{-1}}$ up to a radius of $r\simeq10\,{\rm km}$ and then decreases outward again. This is different to most published grid-based simulations but similar to one simulation published by \citet{shibata06_ns-merger} and recent simulations including a viscous modeling of turbulence \citep{shibata17,radice17}. A plausible explanation of the similarities and dissimilarities of our SPH results and published $\Omega$-profiles may be related to a larger numerical viscosity of our SPH simulations and of the early grid-based simulations. In any case, we stress that this difference is restricted to the inner core, which is anyway stable to the MRI in either way (whether $\Omega$ is flat or increasing). For our analysis of MRI growth, the only relevant region is therefore the outer part of the HMNS, where the frequency decreases outwards. In this region, our late-time rotation profile is qualitatively but also quantitatively in the ballpark of the published results (compare e.g. to Figure 11 of \citet{hanauske16}).
In order to determine the sensitivity of our results to the rotation profile, we introduce a rotation profile alternative to the ones used in Sections~\ref{sec:viscous_MRI}-\ref{sec:sigma_lambda} (equation~(\ref{eq:exponential_fit}) and our late-time profile). This profile (shown by the solid blue line in Figure~\ref{fig:profiles}) has the following functional form:
\begin{equation}
\Omega(r) = \Omega_1\frac{\left(r/r_1\right)^{-3/2}}{1+\left(r/r_{\rm max}\right)^{-3}},
	\label{eq:hanauske_profile}
\end{equation}
with $\Omega_1=2\pi\times10^3\, {\rm s^{-1}}$, $r_1=15\,{\rm km}$, and $r_{\rm max}=7\,{\rm km}$. It approximates the behavior of the models with the highest maximum of $\Omega$ in Figure~11 of \citet{hanauske16} (their models APR4-M125 and APR4-M135). We repeated our analysis of the late-time profile but with this alternative rotation law and show the result in Figure~\ref{fig:MRI_regimes_3}. Compared to Figure~\ref{fig:MRI_regimes_2}, the regime of slow MRI is somewhat less extended due to the larger values of $\Omega$. Overall the differences are nonetheless small and the qualitative picture remains unchanged. Overall, our set of three rotation velocity profiles is representative of the full range of profiles observed in Figure 11 of \citet{hanauske16}: a low-$\Omega$ curve like their model GNH3-M135 is close to our exponential fit at radii larger than $5-10\,{\rm km}$, the profile given by equation~(\ref{eq:hanauske_profile}) is close to high-$\Omega$ profiles (their models APR4-M125 and APR4-M135), and finally our late-time profile lies in-between. Our results demonstrate that the differences in the $\Omega(r)$ behavior do not have a crucial influence on our main conclusions, the differences in the MRI growth rate and wavelength being noticeable but not major overall (compare figures~\ref{fig:MRI_regimes}, \ref{fig:MRI_regimes_2} and \ref{fig:MRI_regimes_3}).

It is more difficult to compare the temperature profiles since the aforementioned papers have not published any azimuthally averaged profiles, but we can make the following discussion based on the color figures of \citet{hanauske16,kastaun16b}. Similarly to the rotation frequency-profile, the late-time temperature profile shows a low-temperature core surrounded by larger temperatures, the maximum temperature being reached at a radius close to the maximum of the rotation frequency. The inner core of these simulations is significantly cooler than ours, but again this difference is irrelevant to our analysis as the inner core is stable to the MRI. On the other hand, the maximum temperature (with values of $T\sim 30\,{\rm MeV}$) seems somewhat larger than ours. If this higher temperature is more realistic than ours, it would only strengthen our result that neutrinos slow down the MRI growth (see Section~\ref{sec:drag_MRI} for a discussion of the impact of the temperature uncertainty).

\subsection{Accretion torus around black hole}
We have also applied the same analysis to models of an accretion torus around a BH,
  which are representative of the remnants expected from NS-BH mergers or from NS-NS mergers leading to prompt or
  delayed BH formation. The numerical models were evolved by \citet{Justbausweinetal2015}, using an M1-type
  neutrino transport scheme \citep{Justobergaulingerjanka2015}, with Newtonian dynamics but
  including a pseudo-Newtonian gravitational potential \citep{Artemovabjoernsson1996}, and by
  employing the $\alpha$-viscosity treatment by \citet{Shakurasunyaev1973} to parametrize turbulent
  angular momentum transport. More details regarding the evolution scheme and model setup can be found in
  \citet{Justbausweinetal2015}. For this study, we consider models M3A8m1a2, M3A8m03a2, and
  M4A8m3a2. The two former models, which were initialized with BH masses of
  $M_{\mathrm{BH}}=3\,M_\odot$ and torus masses of $m_{\mathrm{torus}}=0.1\,M_\odot$ and
  $0.03\,M_\odot$, respectively, are representative for remnants of binary NS
  mergers. Model M4A8m3a2 with $M_{\mathrm{BH}}=4\,M_\odot$ and $m_{\mathrm{torus}}=0.3\,M_\odot$
  represents the remnant of a BH-NS merger. We note that $M_{\mathrm{BH}}$ in this
  model is rather low in view of the expected BH mass distribution
  \citep[e.g.][]{dominik12}. A less massive and thus smaller BH typically
  leads to higher densities and temperatures in the torus. As a consequence, in our model the
  (already small) impact of neutrinos on the MRI tends to be overestimated with respect to more
  likely cases of higher BH mass. For all three models the $\alpha$-parameter for the
  viscosity is 0.02, the BH spin parameter is $0.8$, and the considered snapshot is 20\,ms after the start of the simulations.

\begin{figure}
  \centering
  \includegraphics[width=\columnwidth]{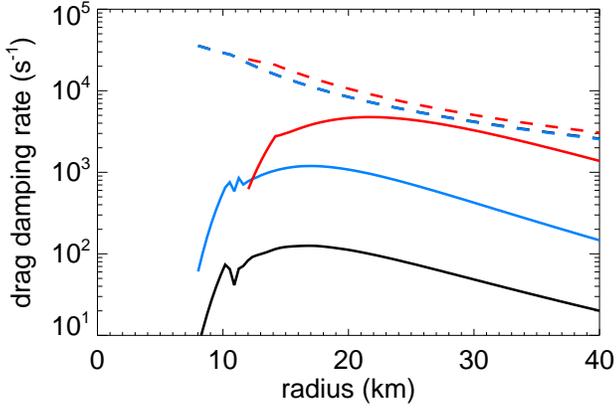}
  \caption{Neutrino drag damping rate in three models of post-merger black-hole accretion tori
      (solid lines). The black and blue lines correspond to tori of masses $0.03\,M_\odot$ and
      $0.1\,M_\odot$, respectively, with a $3\,M_\odot$ BH, while the red lines correspond
      to a torus of mass $0.3\,M_\odot$ with a $4\,M_\odot$ BH. The dashed lines show the
      corresponding angular frequencies (with the black line being nearly identical to the blue
      line).}
  \label{fig:merger}%
\end{figure}

In all three models the neutrino mean free path is comparable to the vertical thickness of the
  torus (i.e. the optical depth is of order unity) and as a result the drag regime is the relevant
  one. We find that the neutrino drag is negligible ($\Gamma\ll\Omega$) in models of low or moderate
  mass of the accretion torus (Figure~\ref{fig:merger}, black and blue lines) and has at most a
  marginal impact ($\Gamma\la\Omega$) in the case of a massive torus (Figure~\ref{fig:merger}, red
  line). This confirms the results of \citet{foucart15}.

\section{Kelvin-Helmholtz instability in the presence of neutrinos}
	\label{sec:KH}
\begin{figure}
\centering
  \includegraphics[width=\columnwidth]{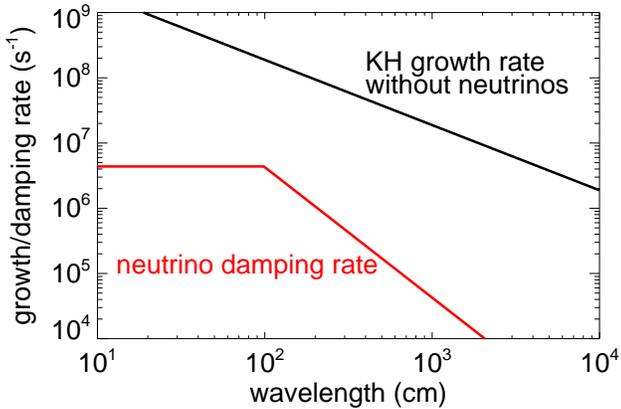}
  \caption{Comparison of the neutrino damping rate (red line) with the growth rate of the
    Kelvin-Helmholtz instability when neutrinos are neglected (black line). The
    Kelvin-Helmholtz instability is not significantly modified by neutrinos since
      the neutrino damping rate is always smaller than the growth rate. The neutrino damping rate
    shown here corresponds to $\rho=10^{14}\,{\rm g\,cm^{-3}}$ and $T=30\,{\rm MeV}$.}
               \label{fig:KH}%
\end{figure}
In this section, we give a rough estimate of the impact of neutrinos on the Kelvin-Helmholtz
instability that develops at the contact interface between the two NSs. To determine
whether neutrinos have a significant impact, we compare the growth rate of the instability (black
line in Figure~\ref{fig:KH}) to the damping rate due to neutrinos (red line in Figure~\ref{fig:KH})
at different wavelengths. The damping rate due to neutrinos is $k^2\nu$ in the viscous regime (where
the viscosity $\nu$ is given by equation~\ref{eq:neutrino_viscosity_approx}), while it is $\Gamma$
as given by equation~\ref{eq:gamma_scaling} in the drag regime. The transition between the two
regimes takes place at a wavelength close to the neutrino mean free path. Initially, the shear layer
is very thin and we will first assume that it can be approximated by a discontinuity with a total
velocity jump $\Delta v \sim 6\times10^9\,{\rm cm\, s^{-1}}$ \citep[see Figure~5 of][]{kiuchi15}. The
growth rate of the Kelvin-Helmholtz instability can then be written as $k\Delta v/2$ (black line in
Figure~\ref{fig:KH}). Figure~\ref{fig:KH} shows, for physical conditions typical of
  NS mergers, that the damping rate due to neutrinos is significantly smaller than the
  growth rate of the Kelvin-Helmholtz instability for all wavelengths. It can be shown (see
  appendix~A) that the criterion for
    neutrinos to have an impact requires a high temperature and a low density, which are
  unlikely to be obtained:
\begin{equation}
\left(\frac{T}{30\,{\rm MeV}}\right)^8 \left( \frac{\rho}{10^{14}\,{\rm g\,cm^{-3}}} \right)^{-2}>25.
	\label{eq:neutrino_KH}
\end{equation}

As time passes, the width $\Delta x$ of the shear layer is expected to increase. The impact of the smoothness of the shear layer on the
instability is to stabilize modes with a wavelength shorter than $\Delta x$. The
  wavelength and growth rate of the fastest growing mode are then a fraction of $\Delta x$ and of
  the maximum vorticity $\sim \Delta v/\Delta x$, respectively, and therefore are close to the
  idealized curve plotted on Figure~\ref{fig:KH}. We therefore expect our conclusion to remain valid
  for a smooth velocity profile. 
  
We conclude that the neutrinos do not change the
growth rate and lengthscale of the Kelvin-Helmholtz instability. This stands in contrast to the
magnetorotational instability, the difference being attributable to the shorter growth timescale of
the Kelvin-Helmholtz instability.

\section{Discussion and conclusion}
	\label{sec:conclusion}
        We have studied the growth of the MRI in compact object mergers, taking into account the
        impact of neutrino viscosity and drag. The main findings can be summarised as follows:
\begin{itemize}
\item Inside the HMNS, the growth rate of the MRI is significantly reduced by neutrino effects
  if the magnetic field strength is lower than $\sim 10^{13}-10^{14}\,{\rm G}$. Depending on the
  radius and magnetic field, the MRI grows in different regimes: the drag regime at wavelengths
  shorter than the neutrino mean free paths for weak magnetic fields ; the viscous regime at
  wavelengths longer than the mean free path for stronger magnetic fields, or a mixed regime at
  intermediate fields (see Figures~\ref{fig:MRI_regimes} and \ref{fig:MRI_regimes_2}). The growth
  timescale becomes longer than $10\,{\rm ms}$ for densities above roughly
    $10^{14}\,{\rm g\,cm^{-3}}$ and magnetic fields weaker than $10^{12}\,{\rm G}$.
\item In the rotationally supported torus around the HMNS, the MRI is at most marginally affected by
  neutrinos. This conclusion also applies to tori in cases of a central BH, thus
    confirming the results of \citet{foucart15}.
\item The growth rate of the Kelvin-Helmholtz instability at the contact interface
    between the two NSs is not significantly influenced by neutrinos.
\end{itemize}

Even though we believe that these results are qualitatively robust, we caution that our analysis suffers from significant quantitative uncertainties. In particular, while we consider an azimuthally averaged angular velocity field, the merger remnant initially shows a more complex flow pattern, which significantly deviates from axisymmetry.

Our analysis demonstrates that inside the HMNS and for initial magnetic fields as expected for typical neutron stars, the MRI is
significantly impacted by neutrinos. We stress that current global numerical simulations, starting
with an unrealistically strong magnetic field, are in a different physical regime of MRI
growth. Results within that regime can therefore not easily be extrapolated to lower initial magnetic fields. 
Given the extreme difficulties in numerically resolving the small length scales at which the MRI grows, local simulations are an indispensable complementary method to study the non-linear phase of the MRI in the presence of neutrinos. The impact of neutrino drag on the non-linear MRI dynamics, for
example, is unknown and could be studied with such an approach.

The slower growth of the MRI that we predict inside the HMNS is likely to lead to a delay of several tens
of ${\rm ms}$ between the collision of binary NSs and the formation of a magnetic field of magnetar strength inside the HMNS. In the scenario where the central engine of a short GRB is a magnetar, we speculate that this could contribute to a delay between the emission of
gravitational waves and the formation of the relativistic jet responsible for the
GRB. Alternatively, the slow MRI growth may imply that another mechanism such as the
Kelvin-Helmholtz instability at the NS-NS collision interface is a more efficient agent during a first phase of magnetic field
amplification. The viability of this mechanism has been debated \citep{obergaulinger10,giacomazzo11,dionysopoulou15} but recent high-resolution
numerical simulations support its efficiency \citep{kiuchi15}. It remains to be
clarified whether the magnetic field can pervade the whole volume of the HMNS and be coherent on
large scales as is implicitly assumed in models of magnetar-powered GRBs.

\section*{Acknowledgements}
At Garching, this work was supported by the Max-Planck--Princeton Center for Plasma Physics (MPPC), by the Deutsche Forschungsgemeinschaft through Excellence Cluster "Universe" (EXC 153), and by the European Research Council through grant ERC-AdG No. 341157-COCO2CASA. AB acknowledges support by the Klaus Tschira Foundation. JG acknowledges support from the European Research Council (grant MagBURSTÐ715368).

\section*{Appendix: Kelvin-Helmholtz instability in the presence of neutrinos}
For the relevant physical conditions, neutrinos can have a significant impact on the Kelvin-Helmholtz instability (i.e. the neutrino damping rate is larger than the growth rate) in an intermediate range of wavelengths between $\lambda_1$ and $\lambda_2$. $\lambda_1$ can be obtained from $k\Delta v/2 = \Gamma$ (drag regime), while $\lambda_2$ follows from $k\Delta v/2=k^2\nu$ (viscous regime). Neutrinos are therefore important for wavenumbers
\begin{equation}
\frac{\Delta v}{2\nu} < k < \frac{2\Gamma}{\Delta v}.
	\label{eq:k_range_KH}
\end{equation}
We can then write the condition for this range to exist and therefore for neutrinos to have an impact on the Kelvin-Helmholtz instability as
\begin{equation}
\frac{\lambda_2}{\lambda_1} = \frac{4\Gamma\nu}{\Delta v^2} > 1.
	\label{eq:neutrino_KH}
\end{equation}
Using equations~\ref{eq:neutrino_viscosity_approx} and~\ref{eq:gamma_scaling}, we find that this condition can be marginally satisfied only for high temperatures and low densities
\begin{equation}
\frac{\lambda_2}{\lambda_1} \simeq 0.04 \left(\frac{T}{30\,{\rm MeV}}\right)^8 \left( \frac{\rho}{10^{14}\,{\rm g\,cm^{-3}}} \right)^{-2}.
	\label{eq:neutrino_KH}
\end{equation}

\bibliography{supernovae}

\bsp
\label{lastpage}

\end{document}